# Extreme value theory applied to the millennial sunspot number series

F.J. Acero[1,2], M.C. Gallego[1,2], J.A. García[1,2], I.G. Usoskin[3], and J.M. Vaquero[2,4]

1 Departamento de Física, Universidad de Extremadura, E-06071 Badajoz, Spain; fjacero@unex.es

2 Instituto Universitario de Investigación del Agua, Cambio Climático y Sostenibilidad (IACYS), Universidad de Extremadura, E-06006 Badajoz, Spain

3 Space Climate Research Unit and Sodankylä Geophysical Observatory, University of Oulu, Finland

4 Departamento de Física, Universidad de Extremadura, E-06800 Mérida, Spain

**Abstract**

In this work, we use two decadal sunspot number series reconstructed from cosmogenic radionuclide data ($^{14}$C in tree trunks, SN-14C and $^{10}$Be in polar ice, SN-10Be) and the Extreme Value Theory to study variability of solar activity during the last 9 millennia. The peaks-over-threshold technique was used to compute, in particular, the shape parameter of the generalized Pareto distribution for different thresholds. Its negative value implies an upper bound of the extreme SN-10Be and SN-14C time series. The return level for 1000 and 10000 years were estimated leading to values lower than the maximum observed values, expected for the 1000-year, but not for the 10000-year return levels, for both series. A comparison of these results with those obtained using the observed sunspot numbers from telescopic observations during the last four centuries suggest that the main characteristics of solar activity have already been recorded in the telescopic period (from 1610 to nowadays) which covers the full range of solar variability from a Grand minimum to a Grand maximum.

**1. Introduction**

The Sun is a variable star, and studies of variations of solar activity may shed light on magnetic activity of cool stars. Solar magnetic activity is observed for the last century as synoptic images of the Sun, and for the last four hundred years in the form of sunspot counts and drawings (see, e.g., review papers by Vaquero 2007; Clette et al. 2014; Hathaway, 2015; Usoskin, 2017). Solar activity depicts a great deal of variability during the last centuries, from the very quiet Maunder minimum in the second half of 17th century (Eddy, 1976; Usoskin et al., 2015) to the modern high-activity episode in the second half of the 20th century (Solanki et al., 2004). Indirect proxies, such as cosmogenic radionuclides in terrestrial archives, can help in reconstructions of solar activity for several millennia in the past (Beer et al. 2012; Usoskin et al. 2017). However, there is still an open question whether the period of the last four hundred years (since 1610 AD) covered by direct solar observations is representative for the entire range of solar variability. In other words, are the observed Maunder minimum and the Modern grand maximum 'typical' for solar activity variations?

In some branches of science, extreme values of significant variables have a special value and meaning. In those cases, one needs to use a statistical theory devoted specifically to an analysis of rare values corresponding to extreme situations. One of the options is the use of the Extreme Value Theory (EVT). In recent years, EVT has been applied in terrestrial (Acero et al. 2014; Wi et al. 2016) and solar climatology (Asensio Ramos 2007; Acero et al. 2017). Extreme values of solar activity have been studied directly using the Sunspot Number series recoded during the last four centuries (Asensio Ramos 2007; Acero et al. 2017) and indirectly using geomagnetic indices compiled during the last recent decades (Siscoe 1976). However, studies using longer time scales have not been performed until now.

Here we analyze reconstructed multi-millennial records of solar activity, based on cosmogenic isotopes, by modern statistical methods, viz. EVT, to evaluate the properties of the extremes of solar activity on the long-time scale.

**2. Data and Methodology**

Two sets of cosmogenic radionuclide data were used as tracers of solar activity (Beer et al. 2012, Usoskin 2017): $^{14}$C in tree trunks and $^{10}$Be in polar ice. The decadal sunspot numbers reconstructed from both radionuclides were considered in this study, as published by Usoskin et al. (2016), and denoted as SN-14C and SN-10Be, respectively. The study period is considered as from 6755 BC through 1645 AD for SN-10Be and from 6755 BC to 1895 AD for SN-14C (Figure 1).

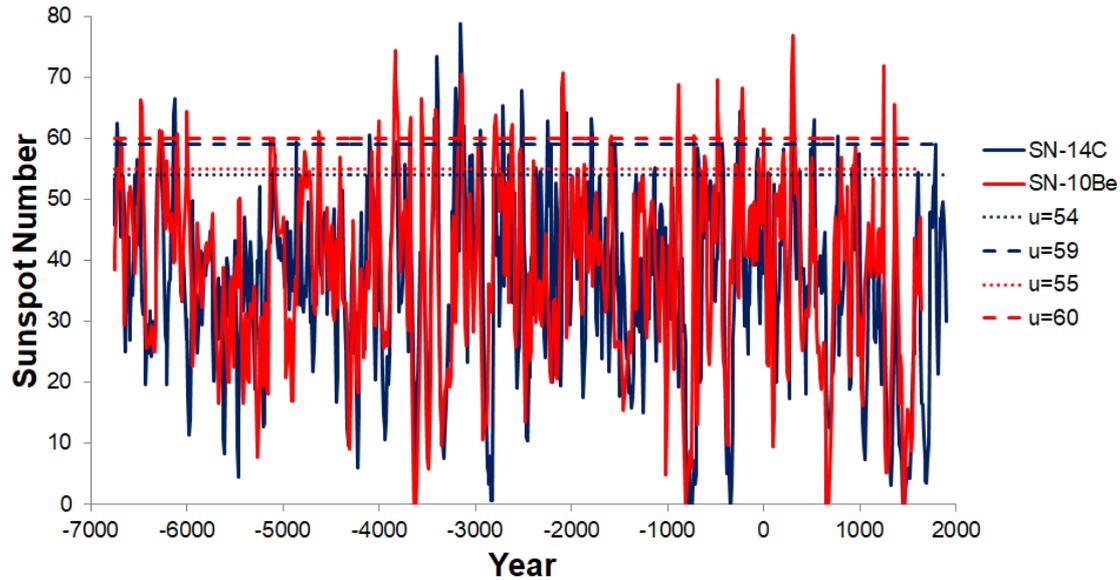

Figure 1. Reconstruction of the decadal sunspot number during the last 9 millennia: (blue) SN-14C and (red) SN-10Be., Different threshold values used in this study are indicated as horizontal lines ($u$ equal to 54 and 59 for SN-14C and 55 and 60 for SN-10Be).

Extreme value theory (EVT) is a branch of statistics dealing with the distribution of excesses, trying to study and quantify the behavior of a process at unusually large or small levels. It seeks to assess, from a given ordered sample of a given variable, the probability of events that are more extreme than any previously observed [Coles 2001], being one of the statistical disciplines most commonly used in the last few decades in many fields such as finance, hydrology and life and earth sciences. EVT is used to study extreme values in a time series and aims to predict the occurrence probability of rare events. Among different approaches used in EVT, we consider the peaks-over-threshold (POT) one. This technique is based on the definition of a high enough threshold and a fit of the exceedances over the threshold to the prescribed statistic.

The POT technique considers all sample values that exceed a predefined upper threshold $u$. The probability distribution of the exceedances over the threshold can be modeled using the generalized Pareto distribution (GPD). The POT approach needs independent observation to avoid too short-range dependences in the time series, when one data point is linked to neighboring ones, and extreme values may cluster together. The cosmogenic radionuclide time series studied in this work show independent observations separated by 10 years. Therefore, it is not necessary to apply a declustering procedure usual in this technique. Both time series were subjected to a GPD analysis. In the asymptotic limit for sufficiently large thresholds, with the observed radionuclide time series SN($t$), the distribution of independent exceedances $X_u(t) = SN(t) - u$ with SN($t$) > $u$ follows a GPD given by

$$P(X < x) = 1 - \left(1 + \frac{\xi \cdot x}{\sigma}\right)^{-1/\xi} \quad (1)$$

with $x > 0$ and $1 + \xi x/\sigma > 0$, where $\sigma$ is the scale parameter, and $\xi$ is the shape parameter ($\xi \neq 0$). Negative shape parameter values indicate that the distribution has an upper bound, while values positive or zero values indicate that the distribution has no upper limit (Coles 2001). The scale parameter $\sigma$ gives information about the variability of the distribution.

In order to apply the above approach to the EVT, it is necessary to choose a correct threshold of extremes from which the exceedances must be evaluated. This threshold $u$ will be not too high in order to the number of exceedances be large enough to minimize the uncertainty of the GPD parameters, but not too low because it would violate the asymptotic basis of the model (Coles 2001). Two methods are available for the threshold selection: the mean residual life plot and the assessment of the stability of the parameter estimates. First, the mean residual life plot was considered, which involves plotting the "mean excess" (the mean value of $X_u(t)$) against $u$, for a range of values of $u$. Such a plot is expected to be linear above the threshold at which the GPD model becomes valid. But, as mentioned by Coles (2001), the interpretation of the mean residual life plot is not always simple in practice. Figure 2 shows the mean residual life plot for both radionuclides (black line) and its confidence interval (dashed lines) based on the approximate normality of sample means (Coles, 2001). Note that there is an approximately linear relationship for the intervals [54, 59] and [55, 60] of the threshold $u$ for SN-14C and SN-10Be, respectively, just when the confidence intervals became larger. For lower values of the threshold $u$, there is another linear zone. However, these lower values of $u$ provide an excessive number of exceedances that do not allow the use of the EVT, as aforementioned. Second, the parameter stability plot involves plotting the parameter estimates from the GPD model against $u$, for a range of values of $u$. These parameters are the shape parameter $\xi$ and the reparametrized scale parameter $\sigma^* = \sigma_u - \xi u$ (see section 4.3.4 of Coles, 2001, for details). The parameter estimates should be stable (i.e. near-constant) above the threshold at which the GPD model becomes valid. The two bottom panels in Figure 2 show the parameter plots and their confidence intervals for SN-14C (left) and SN-10Be (right). This method also confirmed that $u$ in the interval [54, 59] for SN-14C and [55, 60] for SN-10Be time series were optimal thresholds, keeping the shape and scale parameters near-constant. Then, from both methods the thresholds selected were [54, 59] for SN-14C and [55, 60] for SN-10Be which correspond approximately with the values in the interval ranged from the 90th to the 95th percentile for each time series.

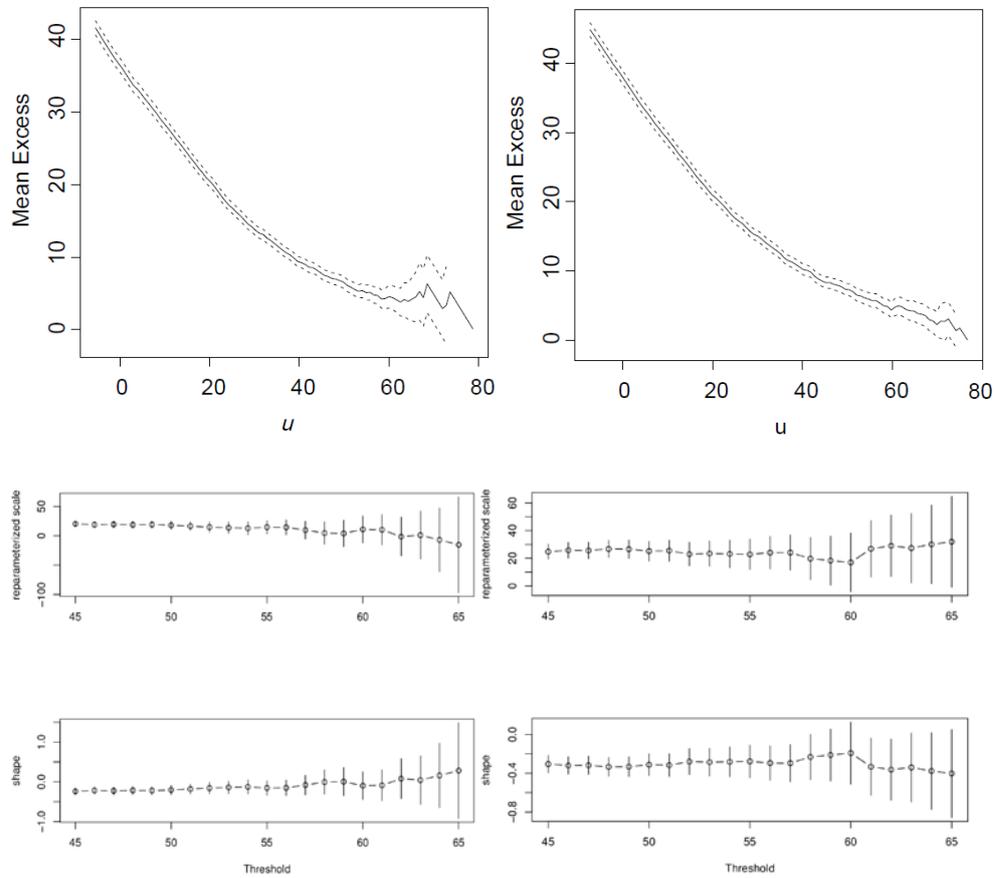

Figure 2. Mean residual life plot (top) and parameter estimates against threshold (two bottom panels) for SN-14C (left) and SN-10Be (right) time series.

Therefore, the independent exceedances for both radionuclides time series were subjected to a GPD analysis. The parameters of the GP distribution in equation (1), were estimated by the maximum likelihood (ML) method using the *in2extRemes* statistical R software package for extreme values (Gilleland & Katz 2013). Once the parameters had been estimated, the confidence interval (CI) for each parameter was evaluated by a bootstrap procedure using 500 replicates (Gilleland & Katz 2013).

Besides, an important property in order to predict the probability of the occurrence of future extreme events is the return level (RL). It is commonly used to convey information about the likelihood of rare events. The *N*-year return level is the level expected to be exceeded once every *N* years, on average, and it was estimated for different values of *N* using both approaches. More details about RL estimations using peaks-over-threshold can be found in Coles (2001). The aforementioned procedure was also used to estimate the RLs and their confidence intervals with the bootstrap procedure.

## 3. Results

Table 1 shows the results for different thresholds analyzed and the corresponding number of exceedances for each threshold, ranging *u* from 54.0 to 58.6 for SN-14C and from 56.0 to 60.4 for SN-10Be. Tables 2 and 3 show the results for the GPD parameters when fitting the data to that distribution for SN-14C and SN-10Be, respectively, and for the different thresholds chosen. These tables list the estimates for the scale and shape parameters and the 95% confidence intervals obtained by bootstrapping. The shape parameter for the different thresholds appears mostly negative even within the 95% CI, implying an upper bound of the extreme values for the SN-10Be and SN-14C radionuclide time series. This result is more relevant for SN-10Be because for all the thresholds considered, except for *u* equal to the 94$^{th}$ percentile, the shape parameter is negative within the 95% CI. For SN-14C, the shape parameter is almost always negative, but the 95% CI sometimes slightly extends to positive values. These results imply, for both radionuclides, that the highest extreme values in relation to solar activity have been reached in the past and are not expected to be exceeded in the future. In order to assess the accuracy of the threshold excess model fitted to each radionuclide time series, different diagnostic plots were used. Figure 3 shows diagnostic plots for the GP fit to the threshold exceedances from the SN-10Be radionuclide time series. The plots confirm the validity of the fitted model: top panel in Figure 3 shows a quantile-quantile (QQ) plot of empirical data quantiles against GP fit quantiles leading to similar distributions with the points lying on the line y=x (solid line). Middle panel in Figure 3 shows a QQ-plot of randomly generated data from the fitted GP against the empirical data quantiles with 95% confidence bands with the points being nearly linear too. Besides, the corresponding density estimate seems consistent between the empirical density of the observed maxima SN with the modelled GP fit density (bottom panel in Figure 3). Figure 4 also shows the diagnostic plot for the GP fit but for the SN-14C time series also validating the fitted model.

Table 1. Estimates of different thresholds and the number of exceedances for both SN-14C and SN-10Be.

|  | SN-14C |  | SN-10Be |  |
| --- | --- | --- | --- | --- |
| Percentile | Threshold ($u$) | Number of exceedances | Threshold ($u$) | Number of exceedances |
| 95th | 58.58 | 44 | 60.42 | 42 |
| 94th | 57.32 | 52 | 59.93 | 51 |
| 93rd | 56.44 | 62 | 58.74 | 59 |
| 92nd | 55.34 | 70 | 57.80 | 68 |
| 91st | 54.64 | 78 | 56.89 | 76 |
| 90th | 54.02 | 87 | 55.97 | 84 |

Table 2. Estimates of the GPD parameters and their 95% confidence intervals (in brackets) obtained by bootstrapping for SN-14C.

|  | SN-14C |
| --- | --- |

| $u$ | Shape parameter ($\xi$) [95% CI] | Scale parameter ($\sigma$) [95% CI] |
|---|---|---|
| 95th | 0.04 [-0.45, 0.32] | 4.01 [2.66, 6.55] |
| 94th | -0.09 [-0.51, 0.14] | 5.09 [3.45, 7.92] |
| 93th | -0.10 [-0.46, 0.13] | 5.32 [3.66, 8.01] |
| 92th | -0.14 [-0.44, 0.04] | 5.95 [4.43, 8.61] |
| 91th | -0.15 [-0.47, 0.01] | 6.14 [4.67, 8.92] |
| 90th | -0.14 [-0.40, 0.03] | 6.14 [4.67, 8.34] |

Table 3. Estimates of the GPD parameters and their 95% confidence intervals (in brackets) obtained by bootstrapping for SN-10Be.

| | SN-10Be | |
|---|---|---|
| $u$ | Shape parameter ($\xi$) [95% CI] | Scale parameter ($\sigma$) [95% CI] |
| 95th | -0.29 [-0.76, -0.06] | 6.31 [4.24, 10.30] |
| 94th | 1.07e-7 [-0.41, 0.26] | 4.16 [2.88, 6.73] |
| 93th | -0.25 [-0.59, -0.07] | 6.21 [4.80, 9.10] |
| 92th | -0.25 [-0.61, -0.08] | 6.50 [5.01, 9.46] |
| 91th | -0.27 [-0.56, -0.12] | 7.01 [5.37, 9.98] |
| 90th | -0.30 [-0.56, -0.15] | 7.59 [5.99, 10.48] |

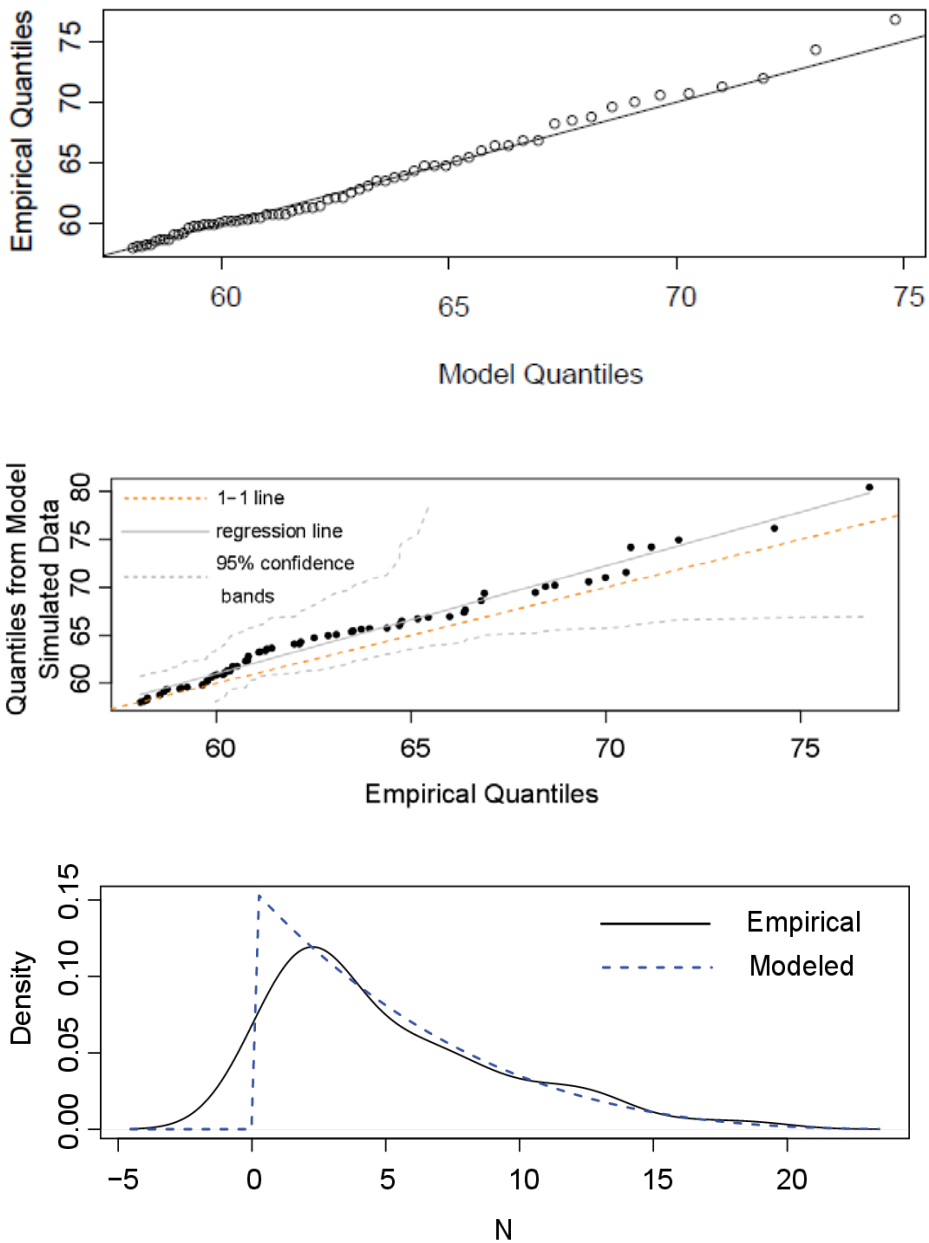

Figure 3. Diagnostic plots from fitting a GPD to the exceedances of SN-10Be. Plots are a QQ-plot of empirical data quantiles against GP fit quantiles (top panel), QQ-plot of randomly generated data from the fitted GP against the empirical data quantiles with 95% confidence bands (middle panel), and (bottom panel) empirical density of the observed maxima SN (solid black line) with GP fit density (dark blue dashed line).

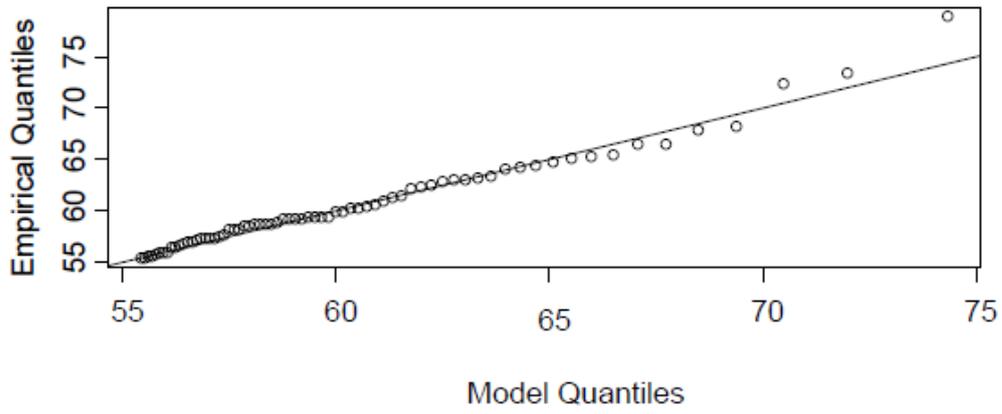

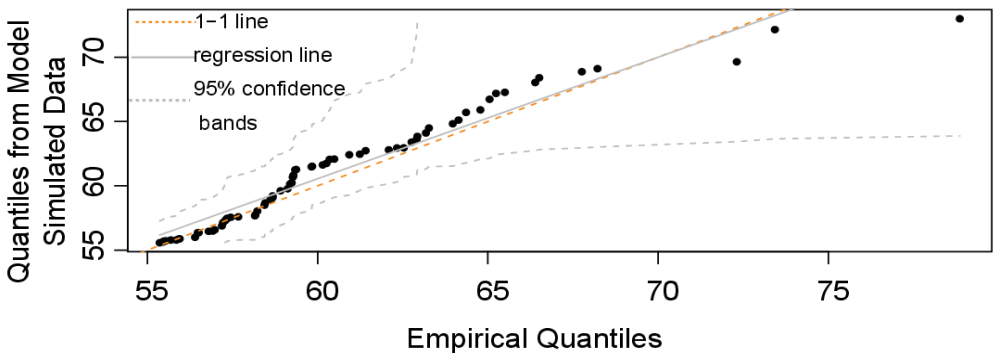

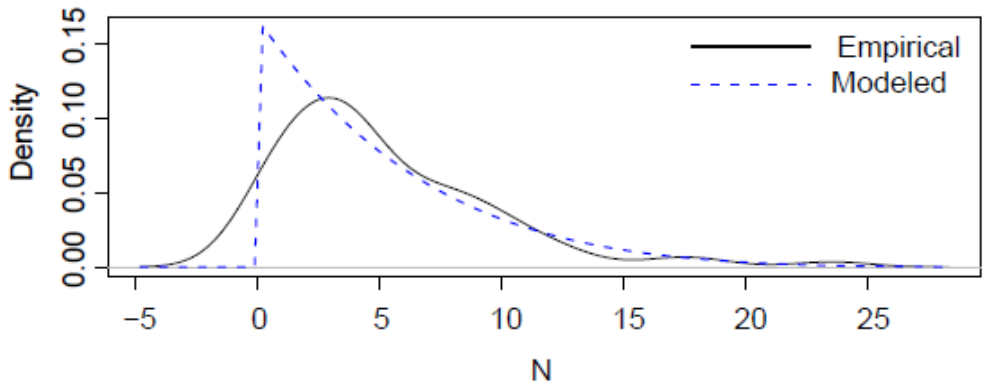

Figure 4. The same as Fig. 3 but for SN-14C.

As mentioned above, a usual procedure for interpreting extreme values uses the return level (RL). As the original time series recorded decadal observations, then the *N*-decadal RL was estimated for *N* = 100, and 1000 decades, corresponding to RLs of 1000 and 10000 years respectively, and the 95% CI was estimated using the bootstrap technique. Table 4 and Table 5 list estimates of the RLs for the two radionuclides time series and for the two values of *N* chosen. The result is not trivial. One would expect that the 1000-year RL lies inside the exceedances range considered for both radionuclides: [54.02, 78.84] for SN-14C and [55.97, 76.77] for SN-10Be for the lowest value of the threshold considered, because the observed time series is longer than 1000 years. However, the results show that even the 10000 year RL corresponding to values of *N* greater than the observed period lie actually also inside the mentioned intervals. For SN-14C and considering the different thresholds, the 10000-year RL varies from 73.9 and 74.5, values lower than the highest observed value 78.8. Similarly for SN-10Be, the 10000-year RL ranges between 74.2 and 76.4, which is also lower than the highest observed value 76.8. The return level plots are shown in Figure 5 for both radionuclides. An extrapolation for the 1000-decade RLs given by the solid line indicates a value lower than the highest observed one, although the maximum for both radionuclides are inside the 95% CI uncertainty of the RL. Therefore, an interesting result is that the RL is reaching a plateau, and it will be unlikely that values greater than the already observed ones will be reached in the future. These results must be considered with some caution, however, because RLs were estimated for a longer period than the observed one.

Table 4. Different estimates of the return level and their 95% confidence intervals (in brackets) obtained by bootstrapping for the SN-14C radionuclide.

|  | SN-14C | |
| --- | --- | --- |
| *u* | 1000-year RL [95% CI] | 10000-year RL [95% CI] |
| 95th | 65.36 [63.50, 67.38] | 74.51 [67.94, 83.35] |
| 94th | 65.78 [63.77, 67.80] | 73.88 [69.01, 80.65] |
| 93th | 65.78 [63.73, 68.02] | 74.26 [69.21, 81.15] |
| 92th | 66.11 [63.99, 68.34] | 73.92 [69.36, 79.62] |
| 91th | 66.03 [64.02, 68.34] | 74.07 [69.57, 79.93] |
| 90th | 66.04 [64.02, 68.46] | 74.07 [69.43, 79.65] |

Table 5. Different estimates of the return level and their 95% confidence intervals (in brackets) obtained by bootstrapping for the SN-10Be radionuclide.

| | SN-10Be | |
|---|---|---|
| $u$ | 1000-year RL [95% CI] | 10000-year RL [95% CI] |
| 95th | 68.58 [66.63, 70.87] | 74.23 [70.80, 77.47] |
| 94th | 67.45 [65.48, 69.82] | 76.44 [70.50, 84.94] |
| 93th | 68.30 [66.31, 70.17] | 74.54 [71.07, 78.16] |
| 92th | 68.29 [66.29, 70.40] | 74.64 [71.10, 78.24] |
| 91th | 68.35 [66.66, 70.25] | 74.39 [70.75, 77.97] |
| 90th | 68.53 [66.89, 70.23] | 74.42 [70.93, 77.53] |

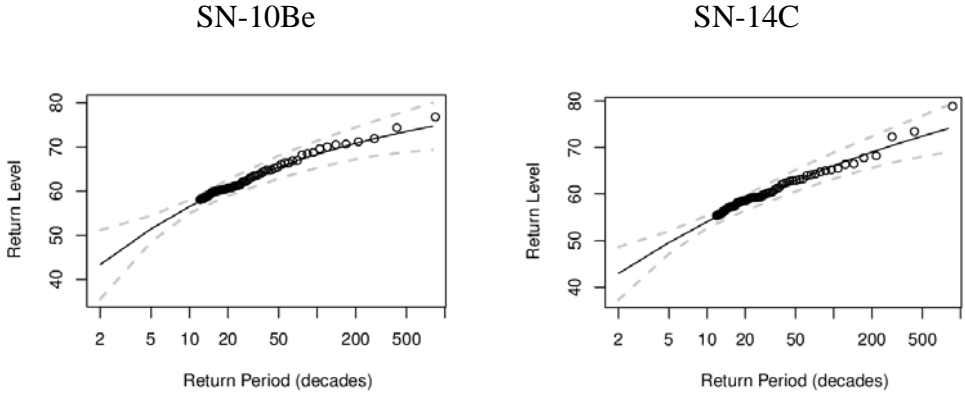

Figure 5. Return level plot of the maximum values for SN-10Be (left) and SN-14C (right) considering as threshold u equal to the 92th percentile, with 95% confidence intervals (dashed lines).

## 4. Conclusions

In this work, we have applied the POT technique, in the context of the Extreme Value Theory, to SN-14C and SN-10Be time series for the last nine millennia. The shape and the scale parameter for the generalized Pareto distribution, which models the probability distribution of the exceedances, were estimated. Different thresholds were considered and validated in order to be accepted as good thresholds in the sense of the EVT. As aforementioned, the information provided by the shape parameter is relevant for an understanding of the behavior of the time series in the sense of extreme values. In this study, the shape parameter was found mostly negative for all the thresholds considered and for both radionuclides, thus revealing the existence of an upper bound for the extremes of the time series. Besides, to interpret the extreme values, the N-year return levels were estimated for two values: N=1000 and 10000 years. As expected, the 1000-year RL lies inside the exceedances range considered for both radionuclides but surprisingly the 10000-year RL also lies in the exceedances range despite of the fact that the observed period is shorter than 10000 years, being probable a higher value. These results suggest that the highest extreme values of the time series for both radionuclides in relation to solar activity have been reached in the past and are not expected to be exceeded in the nearest future. Moreover, the return levels are reaching a plateau, and it will be unlikely that the sunspot numbers will attain values greater than the already observed ones in the future. This generally agrees with a similar analysis carried out using the sunspot number based in telescopic data for the last four centuries (Acero et al. 2017).

Thus, the main characteristics of the solar activity have already been observed in the telescopic period (the last four centuries) which covers the full range of solar variability from a grand minimum to a grand maximum. In any case, this approach and the results are limited by the decadal resolution of the isotope data used in this work.

**Acknowledgements**

Support from the Junta de Extremadura (Research Group Grant GR15137) and from the Ministerio de Economía y Competitividad of the Spanish Government (AYA2014-57556-P) is gratefully acknowledged. I.U. acknowledges support by the Center of Excellence ReSoLVE (Academy of Finland Project No. 272157).